\documentclass[journal=jctcce,manuscript=article]{achemso}

\usepackage{amsmath}
\usepackage{amssymb}
\usepackage{physics}
\usepackage{graphicx} 
\usepackage{enumerate}
\usepackage[inline]{enumitem}
\usepackage{bm}
\usepackage{tcolorbox}
\usepackage{xcolor}
\usepackage[normalem]{ulem}
\usepackage{tikz}

\usepackage{hyperref}
\hypersetup{hidelinks,
  colorlinks=true,
  allcolors=blue,
  pdfstartview=Fit,
  breaklinks=true
}

\newcommand{\im}{\textup{i}}
\newcommand{\Lv}{\mathcal{L}}
\newcommand{\Proj}{\bm{\mathcal{P}}}
\newcommand{\Qroj}{\bm{\mathcal{Q}}}

\author{Rui-Hao Bi}
\affiliation{Department of Chemistry, School of Science and Research Center for Industries of the Future, Westlake University, Hangzhou, Zhejiang 310024, China}

\author{Wei Liu}
\affiliation{Department of Chemistry, School of Science and Research Center for Industries of the Future, Westlake University, Hangzhou, Zhejiang 310024, China}

\author{Wenjie Dou}%
\email{douwenjie@westlake.edu.cn} 
\affiliation{Department of Chemistry, School of Science and Research Center for Industries of the Future, Westlake University, Hangzhou, Zhejiang 310024, China}
\affiliation{Institute of Natural Sciences, Westlake Institute for Advanced Study, Hangzhou, Zhejiang 310024, China}
\affiliation{Key Laboratory for Quantum Materials of Zhejiang Province, Department of Physics, School of Science and Research Center for Industries of the Future, Westlake University, Hangzhou, Zhejiang 310024, China}

\title{Generalized quantum master equation from memory kernel coupling theory}


\begin{document}


\begin{abstract}
The generalized quantum master equation provides a powerful framework for non-Markovian dynamics of open quantum systems. However, the accurate and efficient evaluation of the memory kernel remains a challenge. In this work, we introduce a comprehensive tensorial extension to the Memory Kernel Coupling Theory (MKCT) to overcome this bottleneck. By elevating the original scalar formalism to a tensorial framework, the extended MKCT enables the calculation of general expectation values and cross-correlation functions. We demonstrate the numerical accuracy and efficiency of this method across multiple benchmark systems: capturing transient populations and coherences in the spin-boson model, resolving the excitonic absorption spectrum of the Fenna-Matthews-Olson complex, and simulating charge mobility in one-dimensional lattice models. These successful applications establish the tensorial MKCT as a highly efficient tool for investigating complex dynamics in open quantum systems.
\end{abstract}


\section{Introduction}
Many critical dynamical processes in chemical physics---such as electron and energy transfer \cite{Marcus1993,dou18electronic_friction,dou2020metal_surface}, spectroscopic response \cite{tanimura2012twodspec, ma2015spec,saraceno2023spec}, and quantum transport \cite{troisi2011revmobility,yan_theoretical_2019,jochen2022acc_rev,li2024transport,Bhattacharyya2024mori,jasrasaria2025transport,takahashi2025carrier}---can be fundamentally modeled as open quantum systems\cite{Zwanzig2001text,Yan2005,Breuer2007OQS}. Accurately simulating these systems requires striking a delicate balance between numerical precision and computational scalability. While exact numerical methods \cite{tanimura1989heom,makri1995quapi,jin2008heom,li2020jpcl,wang2022DEOM} provide reliable benchmarks but remain computationally demanding, limiting their applicability to small or medium-sized models.  Conversely, a wide variety of approximate techniques have been developed to handle larger systems, spanning perturbative approaches \cite{redfield1957redfield,Tokuyama1976TCL}, mode-coupling theories \cite{Reichman2002mct1,Reichman2002mct2}, semiclassical methods \cite{Meyer1979MeyerMiller,Stock1997StockThoss,liu2007lscivr}, and mixed quantum-classical dynamics \cite{tully90fssh,Craig2005TSHKSDFT,Wang2015SH_progress,Mannouch2023MASH}. Among these theoretical frameworks, the Generalized Quantum Master Equation (GQME) \cite{Nakajima1958projection,zwanzig1960projection,Zwanzig1961projection,mori1965projection,kelly_generalized_2016,xiang_2021_gqmerev} has emerged as a particularly powerful tool. The GQME becomes highly advantageous when the memory kernel of the system decays much faster than the reduced system dynamics, a regime that is frequently encountered in complex open quantum systems \cite{Cohen2011exact_kernel,andres2025gqme,liu2025mkct,Dan2022kernel_faster}.

In GQMEs (Eq.~\ref{eqn:mori_gqme}), calculating the frequency matrix $\bm{\Omega}$ (the first-order moment) is straightforward, but evaluating the memory kernel $\bm{\mathcal{K}}(t)$ remains a significant challenge. Because brute-force calculations are largely restricted to simple model systems \cite{zhang_2016_directkernel}, realistic applications typically rely on projection-free inputs to approximate $\bm{\mathcal{K}}(t)$ \cite{shi2003memker,shi2004semiclassical,Mulvihill2021kernels_rev,xiang_2021_gqmerev}. Alternatively, non-perturbative diagrammatic approaches based on the influence functional have also been developed \cite{Ivander2024unified}. Furthermore, data-driven techniques---such as the transfer tensor method (TTM) \cite{javier2014TTM}, dynamic mode decomposition \cite{schmid_2010_dmd,wei2023dmd}, and machine learning \cite{xiang2024kernel_nn}---have emerged as powerful tools to extrapolate long-time dynamics from limited short-time snapshots.

Building on these developments, we recently introduced Memory Kernel Coupling Theory (MKCT) \cite{liu2025mkct,bi_2025_universal,liu_2025_dmrgmkct} to efficiently calculate memory kernels. Instead of relying on computationally demanding time propagation, MKCT determines the memory kernel by solving coupled equations of auxiliary higher-order moments and kernels. This approach has been used to faithfully compute the correlation functions of both bosonic \cite{liu2025mkct,bi_2025_universal} and fermionic \cite{liu2025mkct,liu_2025_dmrgmkct} open quantum systems. However, the original scalar formalism is restricted to computing autocorrelation functions, $C_{AA}(t)$. It cannot evaluate cross-correlation functions, $C_{AB}(t)$, or the time evolution of general expectation values, such as state populations and coherences.

In this paper, we show how these limitations can be addressed through a straightforward tensorial extension of the original MKCT. In Sec.~\ref{sec:theory}, we outline this extension, which preserves the equation structure of the scalar formulation while promoting the moments and memory kernels to tensorial quantities. Sec.~\ref{sec:num} presents a comprehensive application of the tensorial MKCT to three representative open quantum systems: the spin-boson model (Sec.~\ref{subsec:tls}), the FMO complex (Sec.~\ref{subsec:spec_fmo}), and two one-dimensional transport models (Sec.~\ref{subsec:transport}). We conclude in Sec.~\ref{sec:conclusion}.

\section{\label{sec:theory}Theory}
\subsection{Preliminary considerations}
Let us consider an open quantum system described by the following Hamiltonian:
\begin{equation}
    \hat{H} = \hat{H}_s + \hat{H}_b + \hat{H}_{sb}.
\end{equation}
Here, $\hat{H}_s$ is the system Hamiltonian, $\hat{H}_b$ is the bath Hamiltonian, and $\hat{H}_{sb}$ is the system-bath interaction Hamiltonian. In the Heisenberg picture, the system operator $\hat{A}$ evolves under the Liouville equation
\begin{equation}
    \dv{}{t} \hat{A}(t) = \frac{\im}{\hbar} \Lv \hat{A}(t) = e^{\im\Lv t} \hat{A},
\end{equation}
where the total Liouvillian $\mathcal{L}$ is given by
\begin{equation}
    \Lv(\cdot)= (\Lv_s + \Lv_b + \Lv_{sb}) (\cdot)=\comm{\hat{H}_s + \hat{H}_b + \hat{H}_{sb}}{(\cdot)}.
\end{equation}

We are interested in the quantum dynamics of an operator $\hat{A}$, characterized by its expectation value and its cross-correlation with another operator $\hat{B}$. Throughout this work, we assume a factorized initial condition $\hat{\rho}_0 = \hat{\sigma}_0 \otimes \hat{\rho}_b^\text{eq}$, where the system is prepared in an arbitrary state $\hat{\sigma}_0$ and the bath is initially in thermal equilibrium: $\hat{\rho}_b^\text{eq} = e^{-\beta\hat{H}_b} / \Tr[e^{-\beta\hat{H}_b}]$. Under this condition, the expectation value evolves as
\begin{equation}\label{eqn:expval}
\expval{\hat{A}(t)} = \Tr[\hat{A}(t) \hat{\rho}_0],
\end{equation}
and the two-time correlation function is given by
\begin{equation}\label{eqn:non_eq_corr}
C_{AB}(t) = \expval{\hat{A}(t)\hat{B}(0)} = \Tr[\hat{A}(t) \hat{B}^{\dagger} \hat{\rho}_0].
\end{equation}
In Eq.~\eqref{eqn:non_eq_corr}, we consider the specific case where the second operator is evaluated at the initial time $t=0$.


We span the system Liouville space using a set of orthonormal basis operators$\{\hat{\phi}_i\}$ that satisfies $\Tr[\hat{\phi}_i\hat{\phi}_j^{\dagger}]=\delta_{ij}$. Using this basis, any operator $\hat{A}$ can be expanded as
\begin{subequations}
\begin{gather}
    \hat{A} = \sum_i \Tr[\hat{A} \hat{\phi}_i^{\dagger}] \hat{\phi}_i = \sum_i A_i \hat{\phi}_i, \\
    \hat{A}^{\dagger} = A_i^* \hat{\phi}_i^{\dagger} = \sum_i \Tr[\hat{\phi}_i \hat{A}^{\dagger}]\hat{\phi}_i^{\dagger}.
\end{gather}
\end{subequations}
By applying this expansion, we can express Eq.~\ref{eqn:expval} as:
\begin{equation}
    \expval{\hat{A}(t)} = \sum_{ij} A_i \mathcal{C}_{ij}(t) \sigma_{0,j}^{*}.
\end{equation}
Here, we have introduced the matrix $\bm{\mathcal{C}}(t)$, whose $i,j$-th element is defined as
\begin{equation}\label{eqn:corr_basis}
    \mathcal{C}_{ij}(t) \equiv \Tr[(e^{\im\Lv t} \hat{\phi}_i)\hat{\phi}_j^{\dagger}\otimes\hat{\rho}_b^\text{eq}], 
\end{equation}
which can be interpreted as the autocorrelation functions between the basis operators \cite{montoya-castillo_approximate_2016,montoya-castillo_approximate_2017}. Furthermore, we note that the definition of $\bm{\mathcal{C}}(t)$ is closely related to the concept of dynamical maps introduced in previous studies \cite{Makri1995dynmaps,javier2014TTM}. Similarly, the correlation function $C_{AB}(t)$ in Eq.~\ref{eqn:non_eq_corr} can also be expressed in $\bm{\mathcal{C}}(t)$ as:
\begin{equation}\label{eqn:cross_corr}
    C_{AB}(t)  = \sum_{ij} A_i C_{ij}(t) \Tr[\hat{\phi}_j (\hat{B} \hat{\sigma}_0)^{\dagger}].
\end{equation}

In the following, we show that $\bm{\mathcal{C}}(t)$ ssatisfies the generalized quantum master equation (GQME) \cite{shi2003memker,kelly_generalized_2016,montoya-castillo_approximate_2016,montoya-castillo_approximate_2017}. To begin, we define the projection operators $\{\Proj_i\}$ as:
\begin{equation}\label{eqn:proj_rules}
    \Proj_i \hat{X} = \hat{\phi}_i (\hat{X}, \hat{\phi}_i) = \hat{\phi}_i \Tr[\hat{X} \hat{\phi}_i^{\dagger} \hat{\rho}_b^{\text{eq}}],
\end{equation}
where $\hat{X}$ is an arbitrary operator in the composite system-bath space, and $(\hat{X}, \hat{\phi}_i)$ denotes the Mori product. By summing the individual projections to form the total projection operator $\Proj \equiv \sum_i \Proj_i$, we define a projection that essentially performs a partial trace over the bath degrees of freedom with respect to the thermal equilibrium state $\hat{\rho}_b^{\text{eq}}$. As detailed in Appendix~\ref{app:mori_matrix}, the operator $\Proj$ leads to the following GQME for the basis autocorrelation functions:
\begin{equation}\label{eqn:mori_gqme}
    \dv{}{t}\bm{\mathcal{C}}(t) = \bm{\Omega} \bm{\mathcal{C}}(t) + \int_{0}^{t} \dd{\tau} \bm{\mathcal{K}}(\tau)\bm{\mathcal{C}}(t-\tau),
\end{equation}
where the frequency matrix $\bm{\Omega}$ and the memory kernel $\bm{\mathcal{K}}$ are defined as
\begin{align}
    \Omega_{ij} &\equiv (\im\Lv \hat{\phi}_i, \hat{\phi}_j),\label{eqn:freq_mat} \\
    \mathcal{K}_{ij}(t) &\equiv (\im\Lv \hat{f}_i(t), \hat{\phi}_j), \label{eqn:kernel}
\end{align}
with the fluctuation operator given by:
\begin{equation}\label{eqn:fluct_op}
    \hat{f}_i(t) = e^{\Qroj\im\Lv t} \Qroj\im\Lv \hat{\phi}_i.
\end{equation}
Here, $\Qroj=\bm{1}-\Proj$ the projection onto the complementary subspace, i.e., the bath degrees of freedom.

\subsection{Memory Kernel Coupling Theory\label{subsec:mkct}}
In the original formulation of MKCT \cite{liu2025mkct}, we considered the scalar autocorrelation $C_{AA}(t)$ of a single operator $\hat{A}$. We found that by defining higher-order moments and memory kernels as:
\begin{subequations}
\begin{gather}
    \Omega_n = ((\im\Lv)^n \hat{A}, \hat{A})(\hat{A}, \hat{A})^{-1}, \\
    K_n(t) = ((\im\Lv)^n \hat{f}(t), \hat{A})(\hat{A}, \hat{A})^{-1}, 
\end{gather}
\end{subequations}
the kernels $K_n(t)$ satisfy a set of initial value problems:
\begin{equation}\label{eqn:mkct_scalar}
    \dv{}{t} K_n(t) = K_{n+1}(t) - K_{1}(t) \Omega_n,
\end{equation}
where the initial conditions are determined by 
\begin{equation}\label{eqn:mkct_scalar_initcond}
    K_n(0) = \Omega_{n+1} - \Omega_n \Omega_1.
\end{equation}
With proper truncation \cite{liu2025mkct,liu2026pmkct}, the MKCT equation (Eq.~\ref{eqn:mkct_scalar}) yields the full hierarchy of higher-order memory kernels $K_n(t)$ using only the information contained in the higher-order moments $\{\Omega_n\}$.

In the following, we demonstrate that the core ideas of the scalar formulation can be generalized to an identical MKCT equation for the tensorial kernel $\bm{\mathcal{K}}(t)$. Specifically, we define the $n$-th order frequency matrix (moment)  $\bm{\Omega}_n$ and memory kernel matrix $\bm{\mathcal{K}}_n(t)$ via their elements:
\begin{gather}
    \Omega_{n,ij} = ((\im\Lv)^n \phi_i, \phi_j), \\
    \mathcal{K}_{n, ij} = ((\im\Lv)^n \hat{f}_i(t), \phi_j).
\end{gather}
Here, the subscript $n, ij$ denotes the $(i,j)$-th element of $n$-th order quantity. Taking the time derivative of $\mathcal{K}_{n,ik}(t)$ yields:
\begin{equation}
\begin{aligned}
    \dv{}{t} \mathcal{K}_{n,ik}(t) 
    &= \left((\im\Lv)^n\left\{\bm{1}-\sum_j \Proj_j\right\}\im\Lv \hat{f}_i(t), \hat{\phi}_j\right), \\
    &= \mathcal{K}_{n+1,ik}(t) - \sum_j K_{1,ij}(t) \Omega_{n, jk},
\end{aligned}
\end{equation}
which simplifies to the matrix equation:
\begin{equation}\label{eqn:mkct_eom}
    \dv{}{t} \bm{\mathcal{K}}_n(t) = \bm{\mathcal{K}}_{n+1}(t) - \bm{K}_1(t)\bm{\Omega}_n.
\end{equation}
At $t=0$, substituting $\hat{f}_i(0) = \Qroj \im\Lv \hat{\phi}_i$ into the matrix MKCT Eq.~\ref{eqn:mkct_eom} yields the initial condition:
\begin{equation}\label{eqn:mkct_init}
    \bm{\mathcal{K}}_n(0) = \bm{\Omega}_{n+1} - \bm{\Omega}_{n}\bm{\Omega}_1.
\end{equation}
Eqs.~\ref{eqn:mkct_eom} and \ref{eqn:mkct_init} parallel the equation structures of the original scalar MKCT formulation [Eqs.~\ref{eqn:mkct_scalar} and \ref{eqn:mkct_scalar_initcond}], except that the moments and memory kernels are now matrix quantities.

Despite their simplicity, the MKCT equations (Eqs.~\ref{eqn:mkct_scalar} and \ref{eqn:mkct_eom}) should be solved carefully. Since simple truncation at the $n$-th order is known to cause numerical instabilities \cite{liu2025mkct,bi_2025_universal,liu2026pmkct}, previous works have introduced two distinct numerical strategies: a dynamic mode decomposition (DMD) approach \cite{schmid_2010_dmd,wei2023dmd} for extrapolating $\mathcal{K}(t)$, and an alternative Pad\'{e} approximant-based method \cite{bi_2025_universal,liu_2025_dmrgmkct} for fitting higher-order terms. A very recently projected MKCT strategy shows promise in systematically converging the scalar MKCT \cite{liu2026pmkct}. In the following, we will extend the Pad\'{e}-based methodology, previously established for scalar kernels, to the tensorial kernels treated in this work.

The Pad\'{e} approach exploits the fact that the memory kernels $\bm{\mathcal{K}}(t)$ decay rapidly. Consequently, a Pad\'{e}-approximant of the series expansion of $\bm{\mathcal{K}}_n(t)$ is sufficient to reproduce the kernel faithfully \cite{bi_2025_universal}. This series expansion is constructed by calculating various orders of the time derivatives of $\bm{\mathcal{K}}_n(t)$ evaluated at $t=0$, which are obtained iteratively using the following equation:
\begin{equation}\label{eqn:rec_K}
\begin{aligned}
    \dv[m]{}{t} \mathcal{K}_{n, ik}(0) 
    &= \left((\im\Lv)^n \left\{\bm{1}-\sum_j \Proj_j\right\} \im\Lv (\Qroj\im\Lv)^{m}\hat{\phi}_i, \hat{\phi}_k\right), \\
    &= \dv[m-1]{}{t} \mathcal{K}_{n+1, ik}(0) - \sum_j \tilde{\Omega}_{m,ij} \Omega_{n, jk}.
\end{aligned}
\end{equation}
Here, we introduce the auxiliary quantity $\tilde{\bm{\Omega}}_m$, defined as
\begin{equation}
    \tilde{\Omega}_{m,ij} = (\im\Lv(\Qroj\im\Lv)^m\hat{\phi}_i, \hat{\phi}_j).
\end{equation}
The auxiliary quantity $\tilde{\bm{\Omega}}_m$ itself satisfies a recursive relation when expanding:
\begin{equation}\label{eqn:rec_O}
\begin{aligned}
    \tilde{\Omega}_{m,ij} 
    &= \left(\im\Lv \left\{\bm{1}-\sum_j \Proj_j\right\} \im\Lv (\Qroj\im\Lv)^{m-1}\hat{\phi}_i, \hat{\phi}_k\right), \\
    &= ((\im\Lv)^2(\Qroj\im\Lv)^{m-1}\hat{\phi}_i, \hat{\phi}_j) - \sum_j \tilde{\Omega}_{m-1,ij} \Omega_{1, jk}.
\end{aligned}
\end{equation}
By iteratively expanding these recursive relations (Eqs.~\ref{eqn:rec_K} and \ref{eqn:rec_O}), we express both $\dv[m]{}{t} \bm{\mathcal{K}}_{n}(0)$ and $\tilde{\bm{\Omega}}_{n}$ so they are exclusively determined by higher-order moments $\{\bm{\Omega}_n\}$:
\begin{gather}
    \dv[m]{}{t} \bm{\mathcal{K}}_{n}(0) = \bm{\Omega}_{m+n+1} - \bm{\Omega}_{m+n}\tilde{\bm{\Omega}}_0 - \dots - \tilde{\bm{\Omega}}_{m} \bm{\Omega}_{n}, \label{eqn:dk}\\
    \tilde{\bm{\Omega}}_n = \bm{\Omega}_{m+1} - \tilde{\bm{\Omega}}_0\bm{\Omega}_m - \dots - \tilde{\bm{\Omega}}_{m-1} \bm{\Omega}_1. \label{eqn:tilde_omega}
\end{gather}
Eqs.~\ref{eqn:dk} and \ref{eqn:tilde_omega} allow us to calculate the time derivatives of $\bm{\mathcal{K}}_{n}(t)$ at $t=0$ up to arbitary orders. By applying Pad\'{e} approximant element-wise, we can obtain an approximation of $\bm{\mathcal{K}}_{n}(t)$. Substituting this result into the hierarchy allows us to terminate the MKCT at the $(n-1)$-th order. 

The tensorial extension of the MKCT successfully addresses some limitations of its scalar predecessor. While the original MKCT formalism cannot be used to predict general expectation values and cross-correlation functions, the tensorial framework yields the complete basis correlation matrix $\bm{\mathcal{C}}(t)$. This comprehensive matrix provides direct access to $\expval{A(t)}$ and $C_{AB}(t)$ via Eqs.~\ref{eqn:expval} and \ref{eqn:cross_corr}, respectively (as demonstrated in Secs.~\ref{subsec:tls} and \ref{subsec:transport}). 

\section{\label{sec:num}Numerical Examples}
In this section, we gauge the performance of the tensorial MKCT method introduced in Sec.~\ref{subsec:mkct}. Unless otherwise specified, we employ an Ohmic spectral density to describe the system-bath interaction:
\begin{equation}\label{eqn:ohmic}
    J(\omega) = \pi \frac{\lambda}{\omega_C} \omega e^{-\omega/\omega_C}
\end{equation}
where $\lambda$ is the reorganization energy and $\omega_C$ is the cutoff frequency. The moments $\bm{\Omega}_n$ for the spin-boson model (Sec.~\ref{subsec:tls}) and tight-binding models (Sec.~\ref{subsec:transport}) are calculated using the exact symbolic method from Ref.~\cite{bi_2025_universal}. For the FMO spectra (Sec.~\ref{subsec:spec_fmo}) and the Holstein model (Sec.~\ref{subsec:transport}), moments are obtained via the Heisenberg-picture dissipaton equation of motion (DEOM) approach  \cite{wang2022DEOM,liu2025mkct}. In the DEOM simulations, the bath correlation functions corresponding to the Ohmic spectral density were decomposed using the time-domain Prony \cite{zihao2022prony} and ESPRIT \cite{taka2024esprit} methods, as implemented in the \texttt{QuTiP 5} package \cite{nori2026qutip}.

\subsection{\label{subsec:tls}Spin-boson model}
Let us consider the spin-boson model, defined by the Hamiltonian:
\begin{equation}
    \hat{H} = \frac{\omega_s}{2} \hat{\sigma}_z + \Omega \hat{\sigma}_x + \hat{\sigma}_z \sum_j c_j \hat{q}_j + \sum_j \frac{\omega_j}{2} (\hat{p}_j^2 + \hat{q}_j^2),
\end{equation}
where a single harmonic bath couples to the system via the $\hat{\sigma}_z$ operator. For this model, the original MKCT was able to accurately calculate the autocorrelation function $C_{AA}(t)$ \cite{liu2025mkct,bi_2025_universal}, but it was incapable of evaluating cross-correlation functions $C_{AB}(t)$ or general expectation values, such as state populations.

These limitations are overcome by the tensorial MKCT introduced in this work. The new approach retains the original procedures: mapping moments ($\bm{\Omega}_n$) to the memory kernel ($\bm{\mathcal{K}}_1(t)$), thereby facilitating the straightforward calculation of the full basis correlation matrix $\bm{C}(t)$ via the GQME (Eq.~\ref{eqn:mori_gqme}). A comparison of each matrix element against exact benchmarks is presented in FIG.~\ref{fig:dynamical_map} of Appendix~\ref{app:tensor_correlation} to demonstrate the numerical fidelity of this extension.

\begin{figure}[htpb]
    \centering
    \includegraphics[width=0.85\linewidth]{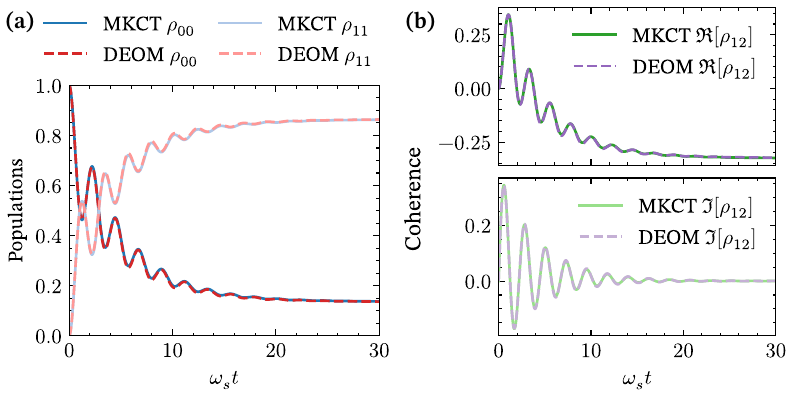}
    \caption{Time evolution of the reduced density matrix for the spin-boson model. Panel (a) depicts the population dynamics, while panel (b) demonstrates the coherence. The simulation parameters are $\omega_s=2$, $\Omega=1$, $\beta=2$, $\lambda=0.2$, and $\omega_\text{C}=5$. A Pad\'{e} approximant of order [9/16] was employed to obtain the memory kernel $\bm{\mathcal{K}}_1(t)$.}
    \label{fig:pop_tls}
\end{figure}

\begin{figure}[htbp]
    \centering
    \includegraphics[width=0.95\linewidth]{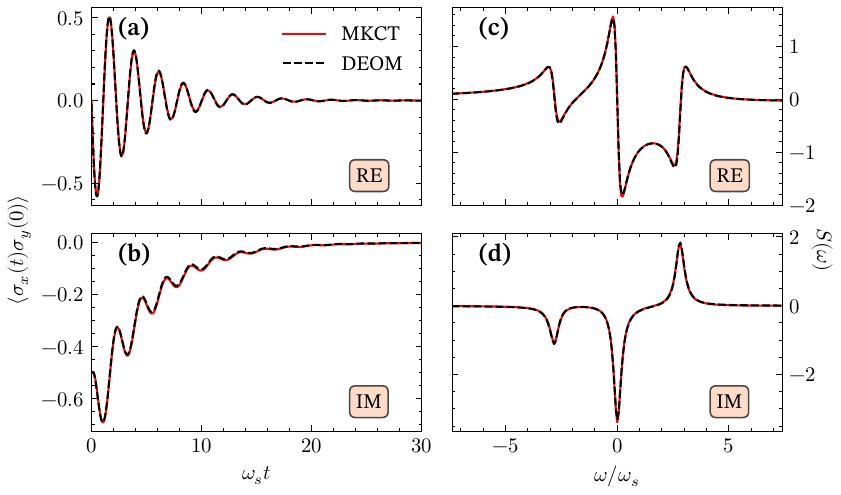}
    \caption{Time evolution of the cross-correlation function $\expval{\sigma_x(t)\sigma_y(0)}$ for a spin-boson model. Panels (a) and (b) depict the real and imaginary parts of the correlation function, respectively. Panels (c) and (d) show the real and imaginary parts of the spectrum $S(\omega)$. The simulation parameters are identical to that of FIG.~\ref{fig:pop_tls}}.
    \label{fig:AB_dynamics}
\end{figure}

Combined with Eq.~\ref{eqn:expval}, $\bm{\mathcal{C}}(t)$ enables us to obtain exact population and coherence dynamics. For example, starting from the initial state $\sigma(0)=\ketbra{0}{0}$, FIG.~\ref{fig:pop_tls} demonstrates that both the populations and coherences obtained from our method agree well with the DEOM results. Furthermore, the same $\bm{\mathcal{C}}(t)$ directly yields the cross-correlation function $C_{\sigma_x\sigma_y}(t)$ via Eq.~\ref{eqn:cross_corr}. As shown in FIG.~\ref{fig:AB_dynamics}, the dynamics generated using this approach are again identical to those from DEOM.

\subsection{\label{subsec:spec_fmo}Absorption spectrum of Fenna–Matthews–Olson (FMO) complex}
The original MKCT was successfully employed to simulate the absorption spectra of two-level systems \cite{liu2025mkct,bi_2025_universal}. However, it has not yet been applied to multi-level systems. To address this, we consider the 7-site FMO complex of \emph{Chlorobium tepidum}, described by the Hamiltonian, partitioned into electronic, vibrational, and coupling terms:
\begin{equation}
    \hat{H}_\text{mol} = \hat{H}_\text{e} + \hat{H}_\text{vib} + \hat{H}_\text{e-vib}.
\end{equation}
The electronic (excitonic) component is given by
\begin{equation}
    \hat{H}_\text{e} = \sum_{m=1}^{7} \epsilon_m \ketbra{m}{m} + \sum_{m=0}^{7}\sum_{n<m} J_{mn}(\ketbra{m}{n} + \ketbra{n}{m}),
\end{equation}
where the site energies $\epsilon_m$ and couplings $J_{mn}$ are taken from Ref.~\cite{cho2005exp_spectra}. The vibrational bath, $\hat{H}_\text{vib}$,  is modeled as a set of independent harmonic oscillators associated with each site:
\begin{equation}
    \hat{H}_\text{vib} = \sum_{m=1}^{7} \sum_{j} \frac{\omega_j}{2} (\hat{p}_{mj}^2 + \hat{q}_{mj}^2).
\end{equation}
The interaction between the excitonic and vibrational degrees of freedom is described by the linear coupling term:
\begin{equation}
    \hat{H}_\text{e-vib} = \sum_{m=1}^{7} \ketbra{m}{m} \sum_j c_{mj} \hat{q}_{mj},
\end{equation}
where the coupling is exclusively diagonal. We assume the spectral density $J_m(\omega)$ is is identical for all sites and follows the Ohmic form defined in Eq.~\ref{eqn:ohmic}. 

To simulate the absorption spectra, we calculate the dipole autocorrelation function as follows:
\begin{equation}
    C_{\bm{\mu}\bm{\mu}}(t) = \Tr[\left(e^{\im\Lv t}\hat{\bm{\mu}}\right)\cdot\hat{\bm{\mu}}\ketbra{0}{0}\otimes\hat{\rho}_b^\text{eq}],
\end{equation}
which is related to the absorption line shape $I(\omega)$  via a Fourier transform:
\begin{equation}
    I(\omega) \propto \Re \int_0^{\infty} \dd{t} C_{\bm{\mu}\bm{\mu}}(t) e^{\im \omega t}.
\end{equation}
We assume that the magnitudes of the transition dipole moments are identical for all seven sites \cite{chen2011fmo,Hein2012fmo}:
\begin{equation}
    \hat{\bm{\mu}} = \sum_{m=1}^{7} \bm{e}_m (\ketbra{0}{m} + \ketbra{m}{0}),
\end{equation}
where the orientation vectors $\bm{e}_m$ of each chromophore are assumed to point from the B to D nitrogen atoms of each Bchl \emph{a} molecule \cite{Tronrud2009structure}.

\begin{figure}[htbp]
    \centering
    \includegraphics[width=0.95\linewidth]{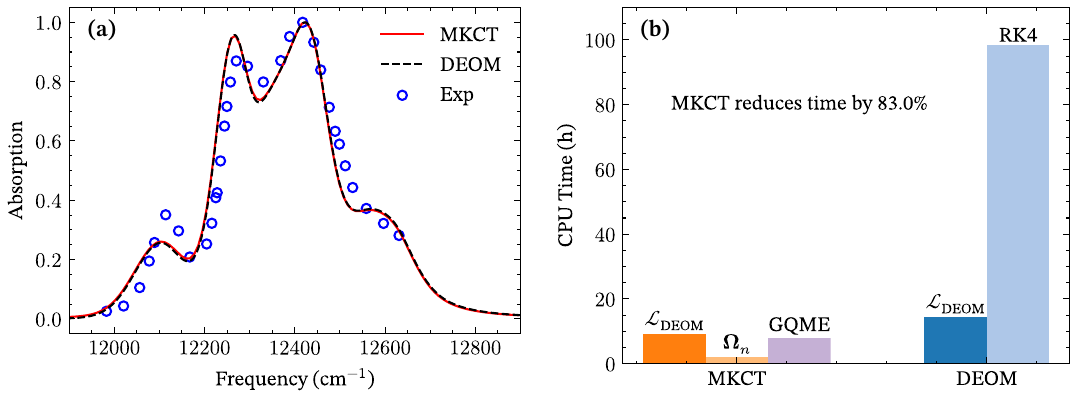}
    \caption{Comparison of the absorption spectrum of the FMO complex computed using MKCT and DEOM. Panel (a) shows the spectra obtained from MKCT (red solid line) and DEOM (black dashed line), alongside the experimental spectrum from Ref.~\cite{cho2005exp_spectra} (black open circles). Panel (b) compares the CPU time required by MKCT and DEOM on an AMD EPYC 7502 processor. The parameter used are $T=77 \, \text{K}$, $\lambda=35 \, \text{cm}^{-1}$, and $\omega_C=50\,\text{fs}^{-1}$. A Pad\'{e} approximant of order [7/13] was used.}
    \label{fig:fmo}
\end{figure}

Following Ref.~\cite{chen2011fmo},  static disorder are incorporated by assuming independent site energy fluctuations with a full-width at half-maximum (FWHM) of $100\,\text{cm}^{-1}$. The final absorption spectra are obtained by averaging 1000 random samples. As demonstrated in FIG.~\ref{fig:fmo}(a), the tensorial MKCT predicts an absorption lineshape identical to the DEOM reference; both results show qualitative agreement with the experimental data. The averaged time-correlation function $C_{\bm{\mu}\bm{\mu}}(t)$ can be found in FIG.~\ref{fig:fmo-time} of Appendix~\ref{app:fmo_timedomain}.

Remarkably, FIG.~\ref{fig:fmo}(b) shows that the MKCT simulation required 80\% less CPU time comparing to DEOM. This efficiency comes from the difference in their numerical approaches: while DEOM requires the propagation of dissipaton density operators until $C_{\bm{\mu}\bm{\mu}}(t)$ decays, MKCT only requires the calculation of a few moments with the DEOM Liouvillian $\mathcal{L}_\text{DEOM}$. These moments  then yield memory kernels via a fast Pad\'{e} approximant, allowing the correlation matrix $\bm{\mathcal{C}}(t)$ to be efficiently computed in the frequency domain via the QME.

\subsection{\label{subsec:transport}Transport properties in one-dimensional chain models}
Thus far, we have demonstrated in Secs.~\ref{subsec:tls} and \ref{subsec:spec_fmo} that the tensorial MKCT framework effectively captures general expectation values and the spectral features of multi-level open quantum systems. To further evaluate the versatility of this method, we now apply it to one-dimensional transport problems. Specifically, we investigate the dissipative Holstein model \cite{li2024transport} and the global-solvent tight-binding model \cite{Lambert2012transport}.

To efficiently simulate the single-mode Holstein model, the original Hamiltonian \cite{HOLSTEIN1959343} is transformed by incorporating the vibrational modes of frequency $\Omega$ into the effective baths, as detailed in Refs.~\cite{Leggett1984,Garg1985,li2024transport}. The resulting effective Hamiltonian is given by
\begin{equation}
    \hat{H}_\text{hol} = \hat{H}_\text{e} + \hat{H}_\text{vib} + \hat{H}_\text{I}.
\end{equation}
Here, the electronic component is described by a one-dimensional tight-binding model:
\begin{equation}\label{eqn:holstein}
    \hat{H}_\text{e} = J \sum_j (\hat{c}_{n+1}^{\dagger}\hat{c}_n + \hat{c}_{n}^{\dagger} \hat{c}_{n+1}),
\end{equation}
where $\hat{c}_n^{\dagger}$ and $\hat{c}_{n}$ are the electronic creation and annihilation operators at the $n$-th site, respectively, and $J$ represents the inter-site hopping integral. Note that the periodic boundary condition is applied. The effective vibrational Hamiltonian is given by
\begin{equation}
    \hat{H}_\text{vib} = \sum_n \sum_j \omega_{nj} \hat{b}_{nj}^{\dagger} \hat{b}_{nj}, 
\end{equation}
where each site is coupled to an independent local bath. Here,
$\hat{b}_{nj}^{\dagger}$ and $\hat{b}_{nj}$ are the bosonic creation and annihilation operators for the $j$-th bath mode at the n-th site.  The system-bath interaction Hamiltonian is defined as
\begin{equation}
    \hat{H}_\text{I} = \sum_n \hat{c}_n^{\dagger}\hat{c}_n \sum_j g_{nj} (\hat{b}_{nj}^{\dagger} + \hat{b}_{nj}).
\end{equation}
where $g_{nj}$ represents the coupling constant to the individual modes. The effect of these local interactions is characterized by the Brownian spectral density function:
\begin{equation}
    J(\omega) = \frac{\lambda\Omega^2\gamma\omega}{(\omega^2 - \Omega^2)^2 + 4\gamma^2\omega^2}.
\end{equation}
In this expression, the dimensionless parameter $\lambda$ characterizes the electron-phonon coupling strength, $\Omega$ is the characteristic vibrational frequency, and $\gamma$ represents the dissipation parameter, or bath friction.

\begin{figure}[htbp]
    \centering
    \includegraphics[width=0.95\linewidth]{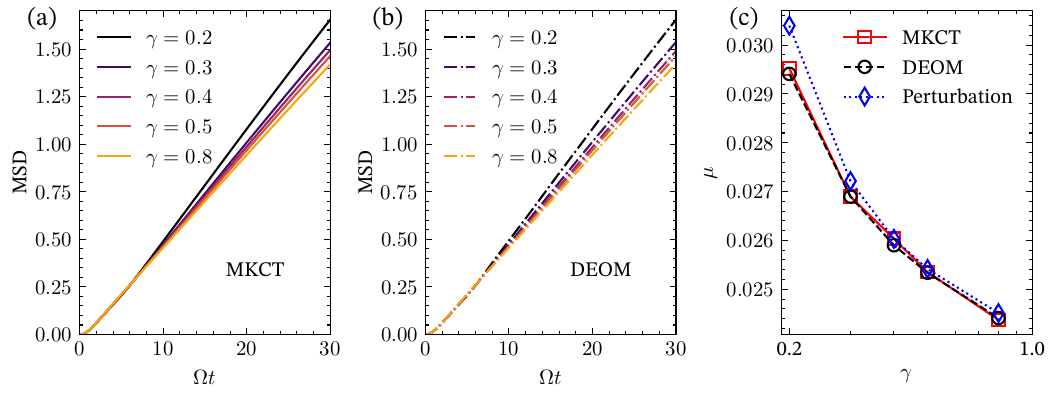}
    \caption{MSD and mobility simulated by MKCT and DEOM at various friction strengths. Panels (a) and (b) show the time evolution of the MSD at different $\gamma$ values computed from MKCT and DEOM, respectively. Panel (c) presents the mobility $\mu$ as a function of the friction coefficient $\gamma$. The simulation parameters are $J = 0.1$, $\lambda = 1$, $\Omega = 1$, and $\beta = 1$.}
    \label{fig:holstein}
\end{figure}

The Holstein model (Eq.~\ref{eqn:holstein}) provides a framework for studying charge mobility. The charge mobility $\mu$ is determined using the Einstein relation: $\mu=\frac{eD}{k_\text{B}T}$. Here, the diffusion constant $D$ is extracted from the long-time limit of the mean square displacement (MSD):
\begin{subequations}
\begin{gather}
    \text{MSD}(t) = \sum_{n}n^2 P_n(t), \\
    \text{D} = \frac{1}{2} \lim_{t\to\infty}\dv{\text{MSD}(t)}{t},
\end{gather}
\end{subequations}
where $P_n(t)$ represents the population at the $n$-th site. In these calculations, we have assumed a unit lattice constant such that the site index $n$ directly corresponds to the spatial coordinate.

As shown in Figs.~\ref{fig:holstein}(a) and \ref{fig:holstein}(b), the time evolution of the MSD calculated via MKCT is in excellent agreement with the DEOM results. Notably, these simulations were performed in the strong electron-phonon coupling regime ($\lambda\Omega>J$),  where approximate methods often fail. Indeed, Fig.~\ref{fig:holstein}(c) demonstrates that perturbation theory \cite{li2024transport} overestimates the mobility in the low-friction limit. In contrast, the tensorial MKCT remains accurate.

Finally, we investigate a tight-binding model coupled to a global solvent, a benchmark system known for its highly temperature-sensitive charge transport dynamics \cite{Lambert2012transport}. The Hamiltonian for is given by
\begin{multline}
    \hat{H}_\text{tb} = \sum_n \epsilon_n \ketbra{n}{n} - V(\ketbra{n}{n+1} + \ketbra{n+1}{n}) + \\
    \sum_j \frac{\hat{p}_j^2}{2m_j}+ \frac{1}{2} m_j \omega_j^2\left(\hat{q}_j - \frac{c_j\hat{s}}{m_j \omega_j^2} \right)^2,
\end{multline}
where $\epsilon_n$ is the energy of the $n$-th site , $V$ is the nearest-neighbor coupling, and $\hat{s}$ is the site displacement operator defined by $\hat{s}\ket{n} = n\ket{n}$. 

We simulate charge transfer originating from a donor site with energy $\epsilon_0 = 1$, while all subsequent acceptor sites are energetically degenerate with $\epsilon_{n>0} = 0$. In the regime $\omega_c > \epsilon_0 > V$, population transfer is strongly bath-mediated, leading to distinct transport mechanisms depending on the bath temperature.

\begin{figure}[htbp]
    \centering
    \includegraphics[width=0.95\linewidth]{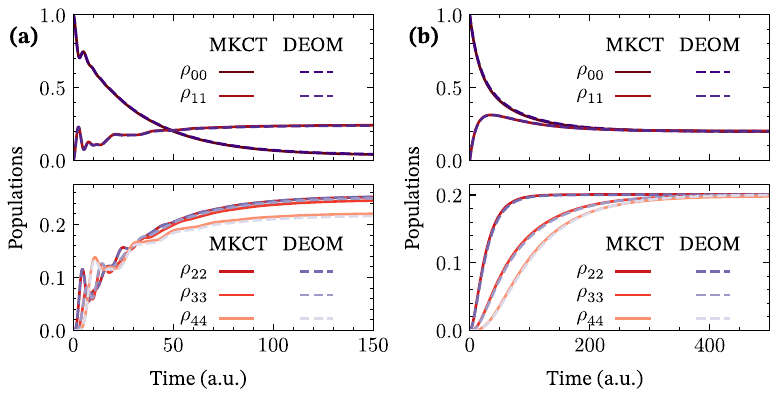}
    \caption{Time evolution of populations for tight-binding model with a global bath. Panel (a) and (b) depicts the results for low temperature ($\beta=2$) and high temeprature ($\beta=0.02$), respectively. The simulation parameters are: $\epsilon_0=1$, $V=0.3$, $\lambda=0.24$, and $\omega_C=4$. A Pad\'{e} approximant of order [7/13] was used.}
    \label{fig:tb}
\end{figure}

FIG.~\ref{fig:tb} illustrates this temperature-dependent behavior with a 5-site model. At low temperatures [Fig.~\ref{fig:tb}(a)], the population transfers almost simultaneously from the donor to all acceptor sites. In contrast, at a higher temperature [panel (b)], site 1 is populated first, followed by sequential transfer down the chain.  Across both of these phenomenologically distinct regimes, the tensorial MKCT shows excellent agreement with the DEOM benchmarks.

\section{\label{sec:conclusion}Conclusion}
In summary, we have introduced a tensorial extension to the MKCT method. This generalized approach maintains an equation structure perfectly parallel to its scalar predecessor, with the crucial distinction that the moments $\bm{\Omega}_n$ and memory kernels $\bm{\mathcal{K}}_n(t)$ are now evaluated as tensorial quantities. The new formulation addresses key limitations of its scalar counterpart, enabling the direct and accurate calculation of general expectation values, including transient state populations and coherences. It also allows for the correct evaluation of correlation functions in multi-level systems at a computational cost significantly lower than that of DEOM. The successful application of this method to one-dimensional transport models demonstrates its robustness across varying temperature and coupling regimes. These results suggest that tensorial MKCT holds promise for simulating complex non-Markovian dynamics in larger open quantum systems.

\begin{acknowledgement}
W.D. acknowledges the support from National Natural Science Foundation of China (No. 22361142829 and No. 22273075) and Zhejiang Provincial Natural Science Foundation (No. XHD24B0301). We thank Westlake university supercomputer center for the facility support and technical assistance.
\end{acknowledgement}


\appendix
\section{\label{app:mori_matrix}Generalized quantum master equation for $\mathbf{\mathcal{C}}(t)$ from Mori-Zwanzig projection techniques.}
The derivation of Eq.~\ref{eqn:mori_gqme} can be found in previous literature (e.g., Ref.~\cite{Zwanzig2001text,montoya-castillo_approximate_2016,xiang_2021_gqmerev}), but we include it here for completeness. We begin with the Liouville–von Neumann equation for the basis operator $\hat{\phi}_i$:
\begin{equation}\label{eqn:lvn_phi}
    \dv{}{t} \hat{\phi}_i(t) = \im\Lv e^{\im\Lv t} \hat{\phi}_i.
\end{equation}
The full propagator $e^{\im\Lv t}$ satisfies the operator identity
\begin{equation}\label{eqn:dyson}
    e^{\im\Lv t} = e^{\Qroj\im\Lv t} + \int_{0}^{t} \dd{\tau} e^{\im\Lv (t - \tau)} \Proj \im \Lv e^{\Qroj\im\Lv\tau}.
\end{equation}
Applying Eq.~\ref{eqn:dyson} to $\Qroj\im\Lv\hat{\phi}_i$ yields:
\begin{align}\label{eqn:op_on_QiLvphi}
    e^{\im\Lv t} (\bm{1}-\sum_j\Proj_j) \im\Lv\hat{\phi}_i = e^{\Qroj\im\Lv t}  \im\Lv\hat{\phi}_i + \int_{0}^{t} \dd{\tau} e^{\im\Lv (t - \tau)} \sum_j \left( \Proj_j \im \Lv e^{\Qroj\im\Lv\tau} \Qroj\im\Lv\hat{\phi}_i \right).
\end{align}
Using Eqs.~\ref{eqn:lvn_phi},\ref{eqn:proj_rules}, and the definitions of the frequency matrix, memory kernel, and fluctuation operator [Eqs.~\ref{eqn:freq_mat}–\ref{eqn:fluct_op}], we obtain:
\begin{equation}\label{eqn:op_basis_evolution}
    \dv{}{t} \hat{\phi}_i(t) = \sum_j \Omega_{ij} \hat{\phi}_j(t) + \int_0^{t} \dd{\tau} \sum_{j} K_{ij}(\tau)\hat{\phi}_j(t-\tau) + \hat{f}_i(t).
\end{equation}
Taking the trace $\Tr[(.)\hat{\phi}_k^{\dagger}\hat{\rho}_b^\text{eq}]$ on both sides of Eq.~\ref{eqn:op_basis_evolution}, we obtain the GQME for the correlation matrix:
\begin{equation}\label{eqn:gqme_before}
    \dv{}{t} \mathcal{C}_{ik} (t) = \sum_j \Omega_{ij} \mathcal{C}_{jk}(t) + \int_0^{t} \dd{\tau} \sum_{j} K_{ij}(\tau)\mathcal{C}_{jk}(t-\tau) + \Tr[\hat{f}_i(t)\hat{\phi}_k^{\dagger} \hat{\rho}_b^\text{eq}].
\end{equation}

To complete the derivation of Eq.~\ref{eqn:mori_gqme}, we show that the term $F_{ik}(t)\equiv\Tr[\hat{f}_i(t)\hat{\phi}_k^{\dagger} \hat{\rho}_b^\text{eq}]$ vanishes for all $t$. We first verify that $F_{ik}(0)=0$:
\begin{equation}\label{eqn:cond1}
\begin{aligned}
    F_{ik}(0)%
    &= \Tr[\left\{(\bm{1} - \sum_j \Proj_j) \im \Lv \hat{\phi}_i \right\}\hat{\phi}_k^{\dagger} \hat{\rho}_b^\text{eq}], \\
    &= \Tr[(\im\Lv\hat{\phi}_i) \hat{\phi}_k^{\dagger} \hat{\rho}_b^\text{eq}] - \sum_j \Tr[(\im\Lv\hat{\phi}_i)\hat{\phi}_j^{\dagger}\hat{\rho}_b^\text{eq}] \Tr[\hat{\phi}_j\hat{\phi}_k^{\dagger}\hat{\rho}_b^\text{eq}] = 0. 
\end{aligned}
\end{equation}
Here, we used the orthonormality condition
\begin{equation}
    \Tr[\hat{\phi}_j\hat{\phi}_k^{\dagger}\hat{\rho}_b^\text{eq}]=\Tr_s[\hat{\phi}_j\hat{\phi}_k^{\dagger}]\Tr_b[\hat{\rho_b^\text{eq}}]=\delta_{jk}.
\end{equation}
Next, we show that $\dv{}{t} F_{ik}(t) = 0$:
\begin{equation}\label{eqn:cond2}
\begin{aligned}
    \dv{}{t} F_{ik}(t) 
    &= \Tr[\left\{(\bm{1} - \sum_j \Proj_j) \im \Lv \hat{f}_i(t) \right\}\hat{\phi}_k^{\dagger}], \\
    &= \Tr[\left\{\im\Lv\hat{f}_i(t)\right\} \hat{\phi}_k^{\dagger} \hat{\rho}_b^\text{eq}] - \sum_j \Tr[\left\{\im\Lv\hat{f}_i(t)\right\}\hat{\phi}_j^{\dagger}\hat{\rho}_b^\text{eq}] \Tr[\hat{\phi}_j\hat{\phi}_k^{\dagger}\hat{\rho}_b^\text{eq}] = 0. 
\end{aligned}
\end{equation}
Since $F_{ik}(0)=0$ and its time derivative vanish, we conclude that $F_{ik}(t)=0$. Substituting this into Eq.~\ref{eqn:gqme_before} yields Eq.~\ref{eqn:mori_gqme} in the main text.

\section{\label{app:tensor_correlation}The basis autocorrelation function of the spin-boson model}
\begin{figure}[htbp]
    \centering
    \includegraphics[width=0.95\linewidth]{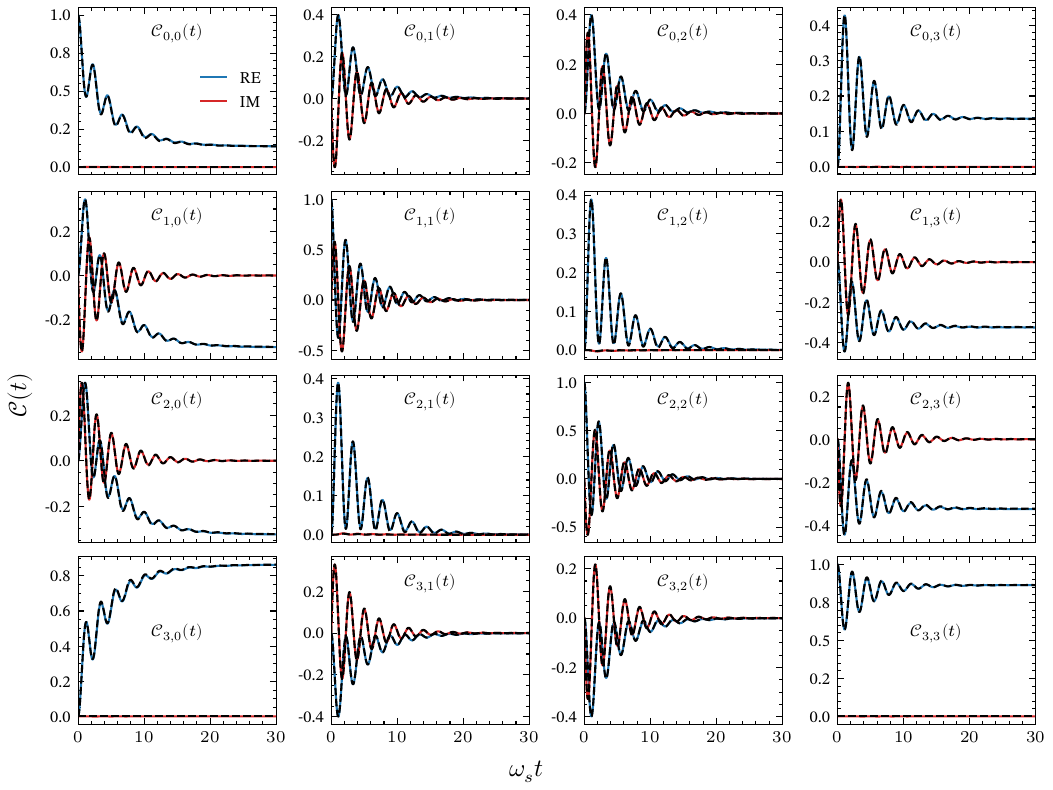}
    \caption{Time evolution of the basis correlation matrix $\bm{\mathcal{C}}(t)$ for the spin-boson model. The solid lines in each subpanel represent the real (blue) and imaginary (red) parts of the upper-triangular matrix elements calculated via MKCT. Reference results from DEOM are shown as dashed black lines. Simulation parameters are identical to those used in Fig.~\ref{fig:pop_tls}.}
    \label{fig:dynamical_map}
\end{figure}

To supplement the population and coherence dynamics discussed in the main text (Sec.~\ref{subsec:tls}), we provide the full time evolution of $\bm{\mathcal{C}}(t)$. FIG.~\ref{fig:dynamical_map} displays the individual matrix elements, comparing the tensorial MKCT results with DEOM benchmarks. 

\section{\label{app:fmo_timedomain}Time-domain correlation function of FMO}

\begin{figure}[htbp]
    \centering
    \includegraphics[width=0.45\linewidth]{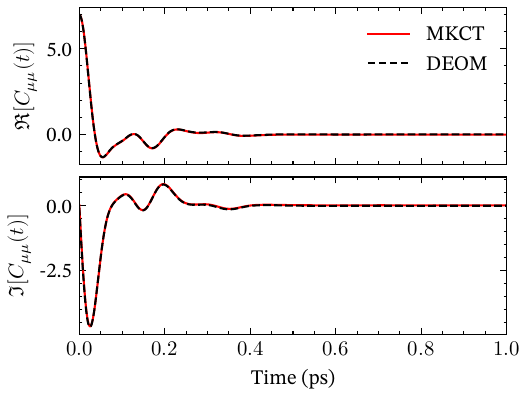}
    \caption{Dipole-dipole correlation function of the FMO complex. The simulation parameters are identical to those used in Fig.~\ref{fig:fmo}}
    \label{fig:fmo-time}
\end{figure}

To complement the absorption spectra of the FMO complex presented in the main text, we plot the time-domain correlation function $C_{\bm{\mu}\bm{\mu}}(t)$ in FIG.~\ref{fig:fmo-time}.

\bibliography{ref}

@article{li2024transport,
	title        = {Is there a finite mobility for the one vibrational mode Holstein model? Implications from real time simulations},
	author       = {Li, Tianchu and Yan, Yaming and Shi, Qiang},
	year         = 2024,
	month        = {03},
	journal      = {The Journal of Chemical Physics},
	volume       = 160,
	number       = 11,
	pages        = 111102,
	doi          = {10.1063/5.0198107},
	issn         = {0021-9606},
	url          = {https://doi.org/10.1063/5.0198107},
	eprint       = {https://pubs.aip.org/aip/jcp/article-pdf/doi/10.1063/5.0198107/19831908/111102\_1\_5.0198107.pdf}
}

@article{ma2015spec,
	title        = {Förster resonance energy transfer, absorption and emission spectra in multichromophoric systems. I. Full cumulant expansions and system-bath entanglement},
	author       = {Ma, Jian and Cao, Jianshu},
	year         = 2015,
	month        = {03},
	journal      = {The Journal of Chemical Physics},
	volume       = 142,
	number       = 9,
	pages        = {094106},
	doi          = {10.1063/1.4908599},
	issn         = {0021-9606},
	url          = {https://doi.org/10.1063/1.4908599},
	eprint       = {https://pubs.aip.org/aip/jcp/article-pdf/doi/10.1063/1.4908599/13811273/094106\_1\_online.pdf}
}

@article{jasrasaria2025transport,
	title        = {Strong anharmonicity dictates ultralow thermal conductivities of type-I clathrates},
	author       = {Jasrasaria, Dipti and Berkelbach, Timothy C.},
	year         = 2025,
	month        = {Jul},
	journal      = {Phys. Rev. B},
	publisher    = {American Physical Society},
	volume       = 112,
	pages        = {014308},
	doi          = {10.1103/s9z1-htzl},
	url          = {https://link.aps.org/doi/10.1103/s9z1-htzl},
	issue        = 1,
	numpages     = 10
}

@book{Breuer2007OQS,
	title        = {The Theory of Open Quantum Systems},
	author       = {Breuer, Heinz-Peter and Petruccione, Francesco},
	year         = 2007,
	month        = {01},
	publisher    = {Oxford University Press},
	doi          = {10.1093/acprof:oso/9780199213900.001.0001},
	isbn         = 9780199213900,
	url          = {https://doi.org/10.1093/acprof:oso/9780199213900.001.0001}
}

@article{tanimura2012twodspec,
	title        = {Reduced hierarchy equations of motion approach with Drude plus Brownian spectral distribution: Probing electron transfer processes by means of two-dimensional correlation spectroscopy},
	author       = {Tanimura, Yoshitaka},
	year         = 2012,
	month        = 11,
	journal      = {The Journal of Chemical Physics},
	volume       = 137,
	number       = 22,
	pages        = {22A550},
	doi          = {10.1063/1.4766931},
	issn         = {0021-9606},
	url          = {https://doi.org/10.1063/1.4766931},
	eprint       = {https://pubs.aip.org/aip/jcp/article-pdf/doi/10.1063/1.4766931/14007229/22a550\_1\_online.pdf}
}

@article{tanimura1989heom,
	title        = {Time Evolution of a Quantum System in Contact with a Nearly Gaussian-Markoffian Noise Bath},
	author       = {Tanimura,  Yoshitaka and Kubo,  Ryogo},
	year         = 1989,
	month        = jan,
	journal      = {Journal of the Physical Society of Japan},
	publisher    = {Physical Society of Japan},
	volume       = 58,
	number       = 1,
	pages        = {101–114},
	doi          = {10.1143/jpsj.58.101},
	issn         = {1347-4073},
	url          = {http://dx.doi.org/10.1143/JPSJ.58.101}
}

@article{makri1995quapi,
	title        = {Numerical path integral techniques for long time dynamics of quantum dissipative systems},
	author       = {Makri, Nancy},
	year         = 1995,
	month        = {05},
	journal      = {Journal of Mathematical Physics},
	volume       = 36,
	number       = 5,
	pages        = {2430--2457},
	doi          = {10.1063/1.531046},
	issn         = {0022-2488},
	url          = {https://doi.org/10.1063/1.531046},
	eprint       = {https://pubs.aip.org/aip/jmp/article-pdf/36/5/2430/19149255/2430\_1\_online.pdf}
}

@article{jin2008heom,
	title        = {Exact dynamics of dissipative electronic systems and quantum transport: Hierarchical equations of motion approach},
	author       = {Jin, Jinshuang and Zheng, Xiao and Yan, YiJing},
	year         = 2008,
	month        = {06},
	journal      = {The Journal of Chemical Physics},
	volume       = 128,
	number       = 23,
	pages        = 234703,
	doi          = {10.1063/1.2938087},
	issn         = {0021-9606},
	url          = {https://doi.org/10.1063/1.2938087},
	eprint       = {https://pubs.aip.org/aip/jcp/article-pdf/doi/10.1063/1.2938087/15413469/234703\_1\_online.pdf}
}

@article{redfield1957redfield,
	title        = {On the Theory of Relaxation Processes},
	author       = {Redfield, A. G.},
	year         = 1957,
	journal      = {IBM Journal of Research and Development},
	volume       = 1,
	number       = 1,
	pages        = {19--31},
	doi          = {10.1147/rd.11.0019}
}

@article{Tokuyama1976TCL,
	title        = {Statistical-Mechanical Theory of the Boltzmann Equation and Fluctuations in µ Space},
	author       = {Tokuyama, Michio and Mori, Hazime},
	year         = 1976,
	month        = 10,
	journal      = {Progress of Theoretical Physics},
	volume       = 56,
	number       = 4,
	pages        = {1073--1092},
	doi          = {10.1143/PTP.56.1073},
	issn         = {0033-068X},
	url          = {https://doi.org/10.1143/PTP.56.1073},
	eprint       = {https://academic.oup.com/ptp/article-pdf/56/4/1073/5358154/56-4-1073.pdf}
}

@article{Reichman2002mct1,
	title        = {A self-consistent mode-coupling theory for dynamical correlations in quantum liquids: Rigorous formulation},
	author       = {Rabani, Eran and Reichman, David R.},
	year         = 2002,
	month        = {04},
	journal      = {The Journal of Chemical Physics},
	volume       = 116,
	number       = 14,
	pages        = {6271--6278},
	doi          = {10.1063/1.1458545},
	issn         = {0021-9606},
	url          = {https://doi.org/10.1063/1.1458545},
	eprint       = {https://pubs.aip.org/aip/jcp/article-pdf/116/14/6271/19307769/6271\_1\_online.pdf}
}

@article{Reichman2002mct2,
	title        = {A self-consistent mode-coupling theory for dynamical correlations in quantum liquids: Application to liquid para-hydrogen},
	author       = {Reichman, David R. and Rabani, Eran},
	year         = 2002,
	month        = {04},
	journal      = {The Journal of Chemical Physics},
	volume       = 116,
	number       = 14,
	pages        = {6279--6285},
	doi          = {10.1063/1.1458546},
	issn         = {0021-9606},
	url          = {https://doi.org/10.1063/1.1458546},
	eprint       = {https://pubs.aip.org/aip/jcp/article-pdf/116/14/6279/19307754/6279\_1\_online.pdf}
}

@article{Meyer1979MeyerMiller,
	title        = {A classical analog for electronic degrees of freedom in nonadiabatic collision processes},
	author       = {Meyer, Hans‐Dieter and Miller, William H.},
	year         = 1979,
	month        = {04},
	journal      = {The Journal of Chemical Physics},
	volume       = 70,
	number       = 7,
	pages        = {3214--3223},
	doi          = {10.1063/1.437910},
	issn         = {0021-9606},
	url          = {https://doi.org/10.1063/1.437910},
	eprint       = {https://pubs.aip.org/aip/jcp/article-pdf/70/7/3214/18916924/3214\_1\_online.pdf}
}

@article{Stock1997StockThoss,
	title        = {Semiclassical Description of Nonadiabatic Quantum Dynamics},
	author       = {Stock, Gerhard and Thoss, Michael},
	year         = 1997,
	month        = {Jan},
	journal      = {Phys. Rev. Lett.},
	publisher    = {American Physical Society},
	volume       = 78,
	pages        = {578--581},
	doi          = {10.1103/PhysRevLett.78.578},
	url          = {https://link.aps.org/doi/10.1103/PhysRevLett.78.578},
	issue        = 4,
	numpages     = {0}
}

@article{liu2007lscivr,
	title        = {Real time correlation function in a single phase space integral beyond the linearized semiclassical initial value representation},
	author       = {Liu, Jian and Miller, William H.},
	year         = 2007,
	month        = {06},
	journal      = {The Journal of Chemical Physics},
	volume       = 126,
	number       = 23,
	pages        = 234110,
	doi          = {10.1063/1.2743023},
	issn         = {0021-9606},
	url          = {https://doi.org/10.1063/1.2743023},
	eprint       = {https://pubs.aip.org/aip/jcp/article-pdf/doi/10.1063/1.2743023/13305336/234110\_1\_online.pdf}
}

@article{tully90fssh,
	title        = {Molecular dynamics with electronic transitions},
	author       = {Tully, John C.},
	year         = 1990,
	month        = {07},
	journal      = {The Journal of Chemical Physics},
	volume       = 93,
	number       = 2,
	pages        = {1061--1071},
	doi          = {10.1063/1.459170},
	issn         = {0021-9606},
	url          = {https://doi.org/10.1063/1.459170},
	eprint       = {https://pubs.aip.org/aip/jcp/article-pdf/93/2/1061/18987588/1061\_1\_online.pdf}
}

@article{Craig2005TSHKSDFT,
	title        = {Trajectory Surface Hopping in the Time-Dependent Kohn-Sham Approach for Electron-Nuclear Dynamics},
	author       = {Craig, Colleen F. and Duncan, Walter R. and Prezhdo, Oleg V.},
	year         = 2005,
	month        = {Oct},
	journal      = {Phys. Rev. Lett.},
	publisher    = {American Physical Society},
	volume       = 95,
	pages        = 163001,
	doi          = {10.1103/PhysRevLett.95.163001},
	url          = {https://link.aps.org/doi/10.1103/PhysRevLett.95.163001},
	issue        = 16,
	numpages     = 4
}

@article{Wang2015SH_progress,
	title        = {Recent Progress in Surface Hopping: 2011–2015},
	author       = {Wang, Linjun and Akimov, Alexey and Prezhdo, Oleg V.},
	year         = 2016,
	journal      = {The Journal of Physical Chemistry Letters},
	volume       = 7,
	number       = 11,
	pages        = {2100--2112},
	doi          = {10.1021/acs.jpclett.6b00710},
	url          = {https://doi.org/10.1021/acs.jpclett.6b00710},
	note         = {PMID: 27171314},
	eprint       = {https://doi.org/10.1021/acs.jpclett.6b00710}
}

@article{Mannouch2023MASH,
	title        = {A mapping approach to surface hopping},
	author       = {Mannouch, Jonathan R. and Richardson, Jeremy O.},
	year         = 2023,
	month        = {03},
	journal      = {The Journal of Chemical Physics},
	volume       = 158,
	number       = 10,
	pages        = 104111,
	doi          = {10.1063/5.0139734},
	issn         = {0021-9606},
	url          = {https://doi.org/10.1063/5.0139734},
	eprint       = {https://pubs.aip.org/aip/jcp/article-pdf/doi/10.1063/5.0139734/19664508/104111\_1\_5.0139734.pdf}
}

@article{Nakajima1958projection,
	title        = {On Quantum Theory of Transport Phenomena: Steady Diffusion},
	author       = {Nakajima, Sadao},
	year         = 1958,
	month        = 12,
	journal      = {Progress of Theoretical Physics},
	volume       = 20,
	number       = 6,
	pages        = {948--959},
	doi          = {10.1143/PTP.20.948},
	issn         = {0033-068X},
	url          = {https://doi.org/10.1143/PTP.20.948},
	eprint       = {https://academic.oup.com/ptp/article-pdf/20/6/948/5440766/20-6-948.pdf}
}

@article{zwanzig1960projection,
	title        = {Ensemble Method in the Theory of Irreversibility},
	author       = {Zwanzig, Robert},
	year         = 1960,
	month        = 11,
	journal      = {The Journal of Chemical Physics},
	volume       = 33,
	number       = 5,
	pages        = {1338--1341},
	doi          = {10.1063/1.1731409},
	issn         = {0021-9606},
	url          = {https://doi.org/10.1063/1.1731409},
	eprint       = {https://pubs.aip.org/aip/jcp/article-pdf/33/5/1338/18820045/1338\_1\_online.pdf}
}

@article{Zwanzig1961projection,
	title        = {Memory Effects in Irreversible Thermodynamics},
	author       = {Zwanzig, Robert},
	year         = 1961,
	month        = {Nov},
	journal      = {Phys. Rev.},
	publisher    = {American Physical Society},
	volume       = 124,
	pages        = {983--992},
	doi          = {10.1103/PhysRev.124.983},
	url          = {https://link.aps.org/doi/10.1103/PhysRev.124.983},
	issue        = 4,
	numpages     = {0}
}

@article{mori1965projection,
	title        = {Transport, Collective Motion, and Brownian Motion},
	author       = {Mori, Hazime},
	year         = 1965,
	month        = {03},
	journal      = {Progress of Theoretical Physics},
	volume       = 33,
	number       = 3,
	pages        = {423--455},
	doi          = {10.1143/PTP.33.423},
	issn         = {0033-068X},
	url          = {https://doi.org/10.1143/PTP.33.423},
	eprint       = {https://academic.oup.com/ptp/article-pdf/33/3/423/5428510/33-3-423.pdf}
}

@article{Mulvihill2021kernels_rev,
	title        = {A Road Map to Various Pathways for Calculating the Memory Kernel of the Generalized Quantum Master Equation},
	author       = {Mulvihill, Ellen and Geva, Eitan},
	year         = 2021,
	journal      = {The Journal of Physical Chemistry B},
	volume       = 125,
	number       = 34,
	pages        = {9834--9852},
	doi          = {10.1021/acs.jpcb.1c05719},
	url          = {https://doi.org/10.1021/acs.jpcb.1c05719},
	note         = {PMID: 34424700},
	eprint       = {https://doi.org/10.1021/acs.jpcb.1c05719}
}

@article{Ivander2024unified,
	title        = {Unified framework for open quantum dynamics with memory},
	author       = {Ivander,  Felix and Lindoy,  Lachlan P. and Lee,  Joonho},
	year         = 2024,
	month        = sep,
	journal      = {Nature Communications},
	publisher    = {Springer Science and Business Media LLC},
	volume       = 15,
	number       = 1,
	doi          = {10.1038/s41467-024-52081-3},
	issn         = {2041-1723},
	url          = {http://dx.doi.org/10.1038/s41467-024-52081-3}
}

@article{shi2003memker,
	title        = {A new approach to calculating the memory kernel of the generalized quantum master equation for an arbitrary system–bath coupling},
	author       = {Shi, Qiang and Geva, Eitan},
	year         = 2003,
	month        = 12,
	journal      = {The Journal of Chemical Physics},
	volume       = 119,
	number       = 23,
	pages        = {12063--12076},
	doi          = {10.1063/1.1624830},
	issn         = {0021-9606},
	url          = {https://doi.org/10.1063/1.1624830},
	eprint       = {https://pubs.aip.org/aip/jcp/article-pdf/119/23/12063/19267201/12063\_1\_online.pdf}
}

@article{Cohen2011exact_kernel,
	title        = {Memory effects in nonequilibrium quantum impurity models},
	author       = {Cohen, Guy and Rabani, Eran},
	year         = 2011,
	month        = {Aug},
	journal      = {Phys. Rev. B},
	publisher    = {American Physical Society},
	volume       = 84,
	pages        = {075150},
	doi          = {10.1103/PhysRevB.84.075150},
	url          = {https://link.aps.org/doi/10.1103/PhysRevB.84.075150},
	issue        = 7,
	numpages     = 5
}

@article{Dan2022kernel_faster,
	title        = {Generalized master equation for charge transport in a molecular junction: Exact memory kernels and their high order expansion},
	author       = {Dan, Xiaohan and Xu, Meng and Yan, Yaming and Shi, Qiang},
	year         = 2022,
	month        = {04},
	journal      = {The Journal of Chemical Physics},
	volume       = 156,
	number       = 13,
	pages        = 134114,
	doi          = {10.1063/5.0086663},
	issn         = {0021-9606},
	url          = {https://doi.org/10.1063/5.0086663},
	eprint       = {https://pubs.aip.org/aip/jcp/article-pdf/doi/10.1063/5.0086663/16539680/134114\_1\_online.pdf}
}

@article{shi2004semiclassical,
	title        = {A semiclassical generalized quantum master equation for an arbitrary system-bath coupling},
	author       = {Shi, Qiang and Geva, Eitan},
	year         = 2004,
	month        = {06},
	journal      = {The Journal of Chemical Physics},
	volume       = 120,
	number       = 22,
	pages        = {10647--10658},
	doi          = {10.1063/1.1738109},
	issn         = {0021-9606},
	url          = {https://doi.org/10.1063/1.1738109},
	eprint       = {https://pubs.aip.org/aip/jcp/article-pdf/120/22/10647/19288987/10647\_1\_online.pdf}
}

@article{kelly_generalized_2016,
	title        = {Generalized quantum master equations in and out of equilibrium: {When} can one win?},
	shorttitle   = {Generalized quantum master equations in and out of equilibrium},
	author       = {Kelly, Aaron and Montoya-Castillo, Andrés and Wang, Lu and Markland, Thomas E.},
	year         = 2016,
	month        = may,
	journal      = {The Journal of Chemical Physics},
	volume       = 144,
	number       = 18,
	pages        = 184105,
	doi          = {10.1063/1.4948612},
	issn         = {0021-9606, 1089-7690},
	url          = {https://pubs.aip.org/jcp/article/144/18/184105/194610/Generalized-quantum-master-equations-in-and-out-of},
	urldate      = {2025-02-24}
}

@article{montoya-castillo_approximate_2017,
	title        = {Approximate but accurate quantum dynamics from the {Mori} formalism. {II}. {Equilibrium} time correlation functions},
	author       = {Montoya-Castillo, Andrés and Reichman, David R.},
	year         = 2017,
	month        = feb,
	journal      = {The Journal of Chemical Physics},
	volume       = 146,
	number       = 8,
	pages        = {084110},
	doi          = {10.1063/1.4975388},
	issn         = {0021-9606, 1089-7690},
	url          = {https://pubs.aip.org/jcp/article/146/8/084110/74277/Approximate-but-accurate-quantum-dynamics-from-the},
	urldate      = {2025-02-24}
}

@article{montoya-castillo_approximate_2016,
	title        = {Approximate but accurate quantum dynamics from the {Mori} formalism: {I}. {Nonequilibrium} dynamics},
	shorttitle   = {Approximate but accurate quantum dynamics from the {Mori} formalism},
	author       = {Montoya-Castillo, Andrés and Reichman, David R.},
	year         = 2016,
	month        = may,
	journal      = {The Journal of Chemical Physics},
	volume       = 144,
	number       = 18,
	pages        = 184104,
	doi          = {10.1063/1.4948408},
	issn         = {0021-9606, 1089-7690},
	url          = {https://pubs.aip.org/jcp/article/144/18/184104/194764/Approximate-but-accurate-quantum-dynamics-from-the},
	urldate      = {2025-02-24}
}

@article{Bhattacharyya2024mori,
	title        = {Mori generalized master equations offer an efficient route to predict and interpret polaron transport},
	author       = {Bhattacharyya, Srijan and Sayer, Thomas and Montoya-Castillo, Andrés},
	year         = 2024,
	journal      = {Chem. Sci.},
	publisher    = {The Royal Society of Chemistry},
	volume       = 15,
	pages        = {16715--16723},
	doi          = {10.1039/D4SC03144J},
	url          = {http://dx.doi.org/10.1039/D4SC03144J},
	issue        = 40
}

@article{yan_theoretical_2019,
	title        = {Theoretical study of charge carrier transport in organic molecular crystals using the {Nakajima}-{Zwanzig}-{Mori} generalized master equation},
	author       = {Yan, Yaming and Xu, Meng and Liu, Yanying and Shi, Qiang},
	year         = 2019,
	month        = jun,
	journal      = {The Journal of Chemical Physics},
	volume       = 150,
	number       = 23,
	pages        = 234101,
	doi          = {10.1063/1.5096214},
	issn         = {0021-9606, 1089-7690},
	url          = {https://pubs.aip.org/jcp/article/150/23/234101/197768/Theoretical-study-of-charge-carrier-transport-in},
	urldate      = {2025-02-24}
}

@article{liu2025mkct,
	title        = {Memory Kernel Coupling Theory: Obtaining Time Correlation Function from Higher-Order Moments},
	author       = {Liu, Wei and Su, Yu and Wang, Yao and Dou, Wenjie},
	year         = 2025,
	month        = {Sep},
	journal      = {Phys. Rev. Lett.},
	publisher    = {American Physical Society},
	volume       = 135,
	pages        = 148001,
	doi          = {10.1103/qvd5-5z6m},
	url          = {https://link.aps.org/doi/10.1103/qvd5-5z6m},
	issue        = 14,
	numpages     = 8
}

@article{saraceno2023spec,
	title        = {First-principles simulation of excitation energy transfer and transient absorption spectroscopy in the CP29 light-harvesting complex},
	author       = {Saraceno, Piermarco and Sláma, Vladislav and Cupellini, Lorenzo},
	year         = 2023,
	month        = 11,
	journal      = {The Journal of Chemical Physics},
	volume       = 159,
	number       = 18,
	pages        = 184112,
	doi          = {10.1063/5.0170295},
	issn         = {0021-9606},
	url          = {https://doi.org/10.1063/5.0170295},
	eprint       = {https://pubs.aip.org/aip/jcp/article-pdf/doi/10.1063/5.0170295/18208511/184112\_1\_5.0170295.pdf}
}

@article{wang2022DEOM,
	title        = {Quantum mechanics of open systems: Dissipaton theories},
	author       = {Wang, Yao and Yan, YiJing},
	year         = 2022,
	month        = 11,
	journal      = {The Journal of Chemical Physics},
	volume       = 157,
	number       = 17,
	pages        = 170901,
	doi          = {10.1063/5.0123999},
	issn         = {0021-9606},
	url          = {https://doi.org/10.1063/5.0123999},
	eprint       = {https://pubs.aip.org/aip/jcp/article-pdf/doi/10.1063/5.0123999/20038347/170901\_1\_5.0123999.pdf}
}

@article{zihao2022prony,
	title        = {Universal time-domain Prony fitting decomposition for optimized hierarchical quantum master equations},
	author       = {Chen, Zi-Hao and Wang, Yao and Zheng, Xiao and Xu, Rui-Xue and Yan, YiJing},
	year         = 2022,
	month        = {06},
	journal      = {The Journal of Chemical Physics},
	volume       = 156,
	number       = 22,
	pages        = 221102,
	doi          = {10.1063/5.0095961},
	issn         = {0021-9606},
	url          = {https://doi.org/10.1063/5.0095961},
	eprint       = {https://pubs.aip.org/aip/jcp/article-pdf/doi/10.1063/5.0095961/16543316/221102\_1\_online.pdf}
}

@article{taka2024esprit,
	title        = {High accuracy exponential decomposition of bath correlation functions for arbitrary and structured spectral     densities: Emerging methodologies and new approaches},
	author       = {Takahashi, Hideaki and Rudge, Samuel and Kaspar, Christoph and Thoss, Michael and Borrelli, Raffaele},
	year         = 2024,
	month        = {05},
	journal      = {The Journal of Chemical Physics},
	volume       = 160,
	number       = 20,
	pages        = 204105,
	doi          = {10.1063/5.0209348},
	issn         = {0021-9606},
	url          = {https://doi.org/10.1063/5.0209348},
	eprint       = {https://pubs.aip.org/aip/jcp/article-pdf/doi/10.1063/5.0209348/19961639/204105\_1\_5.0209348.pdf}
}

@article{nori2026qutip,
	title        = {QuTiP 5: The Quantum Toolbox in Python},
	author       = {Neill Lambert and Eric Giguère and Paul Menczel and Boxi Li and Patrick Hopf and Gerardo Suárez and Marc Gali and  Jake Lishman and Rushiraj Gadhvi and Rochisha Agarwal and Asier Galicia and Nathan Shammah and Paul Nation and J.R.          Johansson and Shahnawaz Ahmed and Simon Cross and Alexander Pitchford and Franco Nori},
	year         = 2026,
	journal      = {Physics Reports},
	volume       = 1153,
	pages        = {1--62},
	doi          = {https://doi.org/10.1016/j.physrep.2025.10.001},
	issn         = {0370-1573},
	url          = {https://www.sciencedirect.com/science/article/pii/S0370157325002704},
	note         = {QuTiP 5: The Quantum Toolbox in Python}
}

@article{javier2014TTM,
	title        = {Non-Markovian Dynamical Maps: Numerical Processing of Open Quantum Trajectories},
	author       = {Cerrillo, Javier and Cao, Jianshu},
	year         = 2014,
	month        = {Mar},
	journal      = {Phys. Rev. Lett.},
	publisher    = {American Physical Society},
	volume       = 112,
	pages        = 110401,
	doi          = {10.1103/PhysRevLett.112.110401},
	url          = {https://link.aps.org/doi/10.1103/PhysRevLett.112.110401},
	issue        = 11,
	numpages     = 5
}

@article{Makri1995dynmaps,
	title        = {Tensor propagator for iterative quantum time evolution of reduced density matrices. I. Theory},
	author       = {Makri, Nancy and Makarov, Dmitrii E.},
	year         = 1995,
	month        = {03},
	journal      = {The Journal of Chemical Physics},
	volume       = 102,
	number       = 11,
	pages        = {4600--4610},
	doi          = {10.1063/1.469508},
	issn         = {0021-9606},
	url          = {https://doi.org/10.1063/1.469508},
	eprint       = {https://pubs.aip.org/aip/jcp/article-pdf/102/11/4600/19014630/4600\_1\_online.pdf}
}

@article{zhang_2016_directkernel,
	title        = {Kinetic Rate Kernels via Hierarchical Liouville-Space Projection Operator Approach},
	author       = {Zhang, Hou-Dao and Yan, YiJing},
	year         = 2016,
	journal      = {The Journal of Physical Chemistry A},
	volume       = 120,
	number       = 19,
	pages        = {3241--3245},
	doi          = {10.1021/acs.jpca.5b11731},
	url          = {https://doi.org/10.1021/acs.jpca.5b11731},
	note         = {PMID: 26757138},
	eprint       = {https://doi.org/10.1021/acs.jpca.5b11731}
}

@article{wei2023dmd,
	title        = {Predicting rate kernels via dynamic mode decomposition},
	author       = {Liu, Wei and Chen, Zi-Hao and Su, Yu and Wang, Yao and Dou, Wenjie},
	year         = 2023,
	month        = 10,
	journal      = {The Journal of Chemical Physics},
	volume       = 159,
	number       = 14,
	pages        = 144110,
	doi          = {10.1063/5.0170512},
	issn         = {0021-9606},
	url          = {https://doi.org/10.1063/5.0170512},
	eprint       = {https://pubs.aip.org/aip/jcp/article-pdf/doi/10.1063/5.0170512/18166311/144110\_1\_5.0170512.pdf}
}

@book{Zwanzig2001text,
	title        = {Nonequilibrium Statistical Mechanics},
	author       = {Zwanzig, Robert W},
	year         = 2001,
	month        = may,
	publisher    = {Oxford University Press},
	address      = {New York, NY}
}

@article{bi_2025_universal,
	title        = {Universal structure of computing moments for exact quantum dynamics: Application to arbitrary system–bath couplings},
	author       = {Bi, Rui-Hao and Liu, Wei and Dou, Wenjie},
	year         = 2025,
	month        = {06},
	journal      = {The Journal of Chemical Physics},
	volume       = 162,
	number       = 22,
	pages        = 224106,
	doi          = {10.1063/5.0273707},
	issn         = {0021-9606},
	url          = {https://doi.org/10.1063/5.0273707},
	eprint       = {https://pubs.aip.org/aip/jcp/article-pdf/doi/10.1063/5.0273707/20553715/224106\_1\_5.0273707.pdf}
}

@misc{liu2026pmkct,
	title        = {Projection-Based Memory Kernel Coupling Theory for Quantum Dynamics: A Stable Framework for Non-Markovian Simulations},
	author       = {Wei Liu and Rui-Hao Bi and Yu Su and Limin Xu and Zhennan Zhou and Yao Wang and Wenjie Dou},
	year         = 2026,
	url          = {https://arxiv.org/abs/2602.10629},
	eprint       = {2602.10629},
	archiveprefix = {arXiv},
	primaryclass = {quant-ph}
}

@article{xiang_2021_gqmerev,
	title        = {Generalized quantum master equation: A tutorial review and recent advances †},
	author       = {Brian, Dominikus and Sun, Xiang},
	year         = 2021,
	month        = 10,
	journal      = {Chinese Journal of Chemical Physics},
	volume       = 34,
	number       = 5,
	pages        = {497--524},
	doi          = {10.1063/1674-0068/cjcp2109157},
	issn         = {1674-0068},
	url          = {https://doi.org/10.1063/1674-0068/cjcp2109157},
	eprint       = {https://pubs.aip.org/cps/cjcp/article-pdf/34/5/497/16739522/497\_1\_online.pdf}
}

@article{liu_2025_dmrgmkct,
	title        = {From higher-order moments to time correlation functions in strongly correlated systems: A DMRG-based memory kernel coupling theory},
	author       = {Liu, Yunhao and Dou, Wenjie},
	year         = 2025,
	month        = 12,
	journal      = {The Journal of Chemical Physics},
	volume       = 163,
	number       = 21,
	pages        = 214102,
	doi          = {10.1063/5.0300474},
	issn         = {0021-9606},
	url          = {https://doi.org/10.1063/5.0300474},
	eprint       = {https://pubs.aip.org/aip/jcp/article-pdf/doi/10.1063/5.0300474/20818576/214102\_1\_5.0300474.pdf}
}

@article{schmid_2010_dmd,
	title        = {Dynamic mode decomposition of numerical and experimental data},
	author       = {Schmid, Peter J.},
	year         = 2010,
	month        = aug,
	journal      = {Journal of Fluid Mechanics},
	volume       = 656,
	pages        = {5--28},
	doi          = {10.1017/S0022112010001217},
	issn         = {0022-1120, 1469-7645},
	url          = {https://www.cambridge.org/core/product/identifier/S0022112010001217/type/journal\_article},
	urldate      = {2026-01-29},
	copyright    = {https://www.cambridge.org/core/terms}
}

@article{cho2005exp_spectra,
	title        = {Exciton Analysis in 2D Electronic Spectroscopy},
	author       = {Cho, Minhaeng and Vaswani, Harsha M. and Brixner, Tobias and Stenger, Jens and Fleming, Graham R.},
	year         = 2005,
	journal      = {The Journal of Physical Chemistry B},
	volume       = 109,
	number       = 21,
	pages        = {10542--10556},
	doi          = {10.1021/jp050788d},
	url          = {https://doi.org/10.1021/jp050788d},
	note         = {PMID: 16852278},
	eprint       = {https://doi.org/10.1021/jp050788d}
}

@article{chen2011fmo,
	title        = {Simulation of the two-dimensional electronic spectra of the Fenna-Matthews-Olson complex using the hierarchical equations of motion method},
	author       = {Chen, Liping and Zheng, Renhui and Jing, Yuanyuan and Shi, Qiang},
	year         = 2011,
	month        = {05},
	journal      = {The Journal of Chemical Physics},
	volume       = 134,
	number       = 19,
	pages        = 194508,
	doi          = {10.1063/1.3589982},
	issn         = {0021-9606},
	url          = {https://doi.org/10.1063/1.3589982},
	eprint       = {https://pubs.aip.org/aip/jcp/article-pdf/doi/10.1063/1.3589982/13818133/194508\_1\_online.pdf}
}

@article{Hein2012fmo,
	title        = {Modelling of oscillations in two-dimensional echo-spectra of the Fenna–Matthews–Olson complex},
	author       = {Hein, Birgit and Kreisbeck, Christoph and Kramer, Tobias and Rodríguez, Mirta},
	year         = 2012,
	month        = {feb},
	journal      = {New Journal of Physics},
	publisher    = {IOP Publishing},
	volume       = 14,
	number       = 2,
	pages        = {023018},
	doi          = {10.1088/1367-2630/14/2/023018},
	url          = {https://doi.org/10.1088/1367-2630/14/2/023018}
}

@article{Tronrud2009structure,
	title        = {The structural basis for the difference in absorbance spectra for the FMO antenna protein from various green sulfur bacteria},
	author       = {Tronrud,  Dale E. and Wen,  Jianzhong and Gay,  Leslie and Blankenship,  Robert E.},
	year         = 2009,
	month        = may,
	journal      = {Photosynthesis Research},
	publisher    = {Springer Science and Business Media LLC},
	volume       = 100,
	number       = 2,
	pages        = {79–87},
	doi          = {10.1007/s11120-009-9430-6},
	issn         = {1573-5079},
	url          = {http://dx.doi.org/10.1007/s11120-009-9430-6}
}

@article{Lambert2012transport,
	title        = {Memory propagator matrix for long-time dissipative charge transfer dynamics},
	author       = {Lambert,  Roberto and Makri,  Nancy},
	year         = 2012,
	month        = aug,
	journal      = {Molecular Physics},
	publisher    = {Informa UK Limited},
	volume       = 110,
	number       = {15–16},
	pages        = {1967–1975},
	doi          = {10.1080/00268976.2012.700408},
	issn         = {1362-3028},
	url          = {http://dx.doi.org/10.1080/00268976.2012.700408}
}

@article{HOLSTEIN1959343,
	title        = {Studies of polaron motion: Part II. The “small” polaron},
	author       = {T. Holstein},
	year         = 1959,
	journal      = {Annals of Physics},
	volume       = 8,
	number       = 3,
	pages        = {343--389},
	doi          = {https://doi.org/10.1016/0003-4916(59)90003-X},
	issn         = {0003-4916},
	url          = {https://www.sciencedirect.com/science/article/pii/000349165990003X}
}

@article{Leggett1984,
	title        = {Quantum tunneling in the presence of an arbitrary linear dissipation mechanism},
	author       = {Leggett,  A. J.},
	year         = 1984,
	month        = aug,
	journal      = {Physical Review B},
	publisher    = {American Physical Society (APS)},
	volume       = 30,
	number       = 3,
	pages        = {1208–1218},
	doi          = {10.1103/physrevb.30.1208},
	issn         = {0163-1829},
	url          = {http://dx.doi.org/10.1103/physrevb.30.1208}
}

@article{Garg1985,
	title        = {Effect of friction on electron transfer in biomolecules},
	author       = {Garg,  Anupam and Onuchic,  José Nelson and Ambegaokar,  Vinay},
	year         = 1985,
	month        = nov,
	journal      = {The Journal of Chemical Physics},
	publisher    = {AIP Publishing},
	volume       = 83,
	number       = 9,
	pages        = {4491–4503},
	doi          = {10.1063/1.449017},
	issn         = {1089-7690},
	url          = {http://dx.doi.org/10.1063/1.449017}
}

@article{xiang2024kernel_nn,
	title        = {How Sophisticated Are Neural Networks Needed to Predict Long-Term Nonadiabatic Dynamics?},
	author       = {Zeng, Hao and Kou, Yitian and Sun, Xiang},
	year         = 2024,
	journal      = {Journal of Chemical Theory and Computation},
	volume       = 20,
	number       = 22,
	pages        = {9832--9848},
	doi          = {10.1021/acs.jctc.4c01223},
	url          = {https://doi.org/10.1021/acs.jctc.4c01223},
	note         = {PMID: 39540684},
	eprint       = {https://doi.org/10.1021/acs.jctc.4c01223}
}

@article{dou18electronic_friction,
	title        = {Perspective: How to understand electronic friction},
	author       = {Dou, Wenjie and Subotnik, Joseph E.},
	year         = 2018,
	month        = {06},
	journal      = {The Journal of Chemical Physics},
	volume       = 148,
	number       = 23,
	pages        = 230901,
	doi          = {10.1063/1.5035412},
	issn         = {0021-9606},
	url          = {https://doi.org/10.1063/1.5035412},
	eprint       = {https://pubs.aip.org/aip/jcp/article-pdf/doi/10.1063/1.5035412/19986259/230901\_1\_1.5035412.pdf}
}

@article{dou2020metal_surface,
	title        = {Nonadiabatic Molecular Dynamics at Metal Surfaces},
	author       = {Dou, Wenjie and Subotnik, Joseph E.},
	year         = 2020,
	journal      = {The Journal of Physical Chemistry A},
	volume       = 124,
	number       = 5,
	pages        = {757--771},
	doi          = {10.1021/acs.jpca.9b10698},
	url          = {https://doi.org/10.1021/acs.jpca.9b10698},
	note         = {PMID: 31916769},
	eprint       = {https://doi.org/10.1021/acs.jpca.9b10698}
}

@article{troisi2011revmobility,
	title        = {Charge transport in high mobility molecular semiconductors: classical models and new theories},
	author       = {Troisi,  Alessandro},
	year         = 2011,
	journal      = {Chemical Society Reviews},
	publisher    = {Royal Society of Chemistry (RSC)},
	volume       = 40,
	number       = 5,
	pages        = 2347,
	doi          = {10.1039/C0CS00198H},
	issn         = {1460-4744},
	url          = {http://dx.doi.org/10.1039/C0CS00198H}
}

@article{jochen2022acc_rev,
	title        = {Charge Transport in Organic Semiconductors: The Perspective from Nonadiabatic Molecular Dynamics},
	author       = {Giannini, Samuele and Blumberger, Jochen},
	year         = 2022,
	journal      = {Accounts of Chemical Research},
	volume       = 55,
	number       = 6,
	pages        = {819--830},
	doi          = {10.1021/acs.accounts.1c00675},
	url          = {https://doi.org/10.1021/acs.accounts.1c00675},
	note         = {PMID: 35196456},
	eprint       = {https://doi.org/10.1021/acs.accounts.1c00675}
}

@article{Marcus1993,
	title        = {Electron Transfer Reactions in Chemistry: Theory and Experiment (Nobel Lecture)},
	author       = {Marcus,  Rudolph A.},
	year         = 1993,
	month        = aug,
	journal      = {Angewandte Chemie International Edition in English},
	publisher    = {Wiley},
	volume       = 32,
	number       = 8,
	pages        = {1111–1121},
	doi          = {10.1002/anie.199311113},
	issn         = {0570-0833},
	url          = {http://dx.doi.org/10.1002/anie.199311113}
}

@article{li2020jpcl,
	title        = {Finite-Temperature TD-DMRG for the Carrier Mobility of Organic Semiconductors},
	author       = {Li, Weitang and Ren, Jiajun and Shuai, Zhigang},
	year         = 2020,
	journal      = {The Journal of Physical Chemistry Letters},
	volume       = 11,
	number       = 13,
	pages        = {4930--4936},
	doi          = {10.1021/acs.jpclett.0c01072},
	url          = {https://doi.org/10.1021/acs.jpclett.0c01072},
	note         = {PMID: 32492339},
	eprint       = {https://doi.org/10.1021/acs.jpclett.0c01072}
}

@article{andres2025gqme,
	title        = {Space-local memory in generalized master equations: Reaching the thermodynamic limit for the cost of a small lattice simulation},
	author       = {Bhattacharyya, Srijan and Sayer, Thomas and Montoya-Castillo, Andrés},
	year         = 2025,
	month        = {03},
	journal      = {The Journal of Chemical Physics},
	volume       = 162,
	number       = 9,
	pages        = {091102},
	doi          = {10.1063/5.0249145},
	issn         = {0021-9606},
	url          = {https://doi.org/10.1063/5.0249145},
	eprint       = {https://pubs.aip.org/aip/jcp/article-pdf/doi/10.1063/5.0249145/20424402/091102\_1\_5.0249145.pdf}
}

@article{Yan2005,
	title        = {QUANTUM MECHANICS OF DISSIPATIVE SYSTEMS},
	author       = {Yan,  YiJing and Xu,  RuiXue},
	year         = 2005,
	month        = may,
	journal      = {Annual Review of Physical Chemistry},
	publisher    = {Annual Reviews},
	volume       = 56,
	number       = 1,
	pages        = {187–219},
	doi          = {10.1146/annurev.physchem.55.091602.094425},
	issn         = {1545-1593},
	url          = {http://dx.doi.org/10.1146/annurev.physchem.55.091602.094425}
}

@article{takahashi2025carrier,
    author = {Takahashi, Hideaki and Borrelli, Raffaele},
    title = {Carrier mobility in Holstein–Peierls models of organic materials: A tensor-train HEOM approach},
    journal = {The Journal of Chemical Physics},
    volume = {163},
    number = {19},
    pages = {194105},
    year = {2025},
    month = {11},
    issn = {0021-9606},
    doi = {10.1063/5.0300292},
    url = {https://doi.org/10.1063/5.0300292},
    eprint = {https://pubs.aip.org/aip/jcp/article-pdf/doi/10.1063/5.0300292/20804276/194105\_1\_5.0300292.pdf},
}

\end{document}